# Compressive Sensing of Color Images Using Nonlocal Higher Order Dictionary

Khanh Quoc Dinh, Thuong Nguyen Canh, and Byeungwoo Jeon

***Abstract*— This paper addresses an ill-posed problem of recovering a color image from its compressively sensed measurement data. Differently from the typical 1D vector-based approach of the state-of-the-art methods, we exploit the nonlocal similarities inherently existing in images by treating each patch of a color image as a 3D tensor consisting of not only horizontal and vertical but also spectral dimensions. A group of nonlocal similar patches form a 4D tensor for which a nonlocal higher order dictionary is learned via higher order singular value decomposition. The multiple sub-dictionaries contained in the higher order dictionary decorrelate the group in each corresponding dimension, thus help the detail of color images to be reconstructed better. Furthermore, we promote sparsity of the final solution using a sparsity regularization based on a weight tensor. It can distinguish those coefficients of the sparse representation generated by the higher order dictionary which are expected to have large magnitude from the others in the optimization. Accordingly, in the iterative solution, it acts like a weighting process which is designed by approximating the minimum mean squared error filter for more faithful recovery. Experimental results confirm improvement by the proposed method over the state-of-the-art ones.**

## I. Introduction

Compressive sensing (CS) [1]–[3] is an emerging technique that samples a signal in a rate much lower than the conventional Nyquist/Shannon sampling rate [4][5] for sparse signals. The possible reduction of sampling rate is attractive for imaging applications such as radar imaging [6], spectral imaging [7], lensless camera [8], flat camera [9], single-pixel imaging [10], and so on, to name a few. Faithful recovery of CS heavily depends on its recovery method from the CS measurements. While many CS recovery methods [11]–[15] have focused on gray images, only few have worked on color images [16]–[18], despite of the fact that color images make more practical sense from user experience viewpoint. Since CS basically assumes signal sparsity in some sparsifying transform domains, its performance can be improved by carefully selecting a sparsifying basis (e.g., either a fixed transformation or a dictionary) and properly regularizing sparsity of the sparse representation corresponding to the selected sparsifying basis.

The sparsifying basis decorrelates the signal of interest and generates its sparse representation. For image signals, a pixel in a color channel is similar typically to its spatially adjacent pixels (spatial correlation), pixels in its other color channels (spectral correlation), and pixels in other similar patches that are even not nearby (nonlocal correlation). One possibility to decorrelate those four dimensional similarities within an image signal is to process those similar pixels as a high dimensional tensor using the recent advancement in the higher order singular value decomposition (HOSVD) [19]. HOSVD can be considered as a generalization of singular value decomposition of a matrix [20] which decomposes correlation along all dimensions of a tensor. This possibility of HOSVD has emerged for some applications such as handwritten digit recognition [19], image recognition [21], image fusion [22], image inpainting [23], and color image/video denoising [24], [25]. Inspired by the excellent decorrelation property of HOSVD, this paper employs it as a tool to adaptively learn a higher order dictionary for each group of (nonlocal) similar patches, in which a patch is processed as a 3D tensor (and then a group of patches is processed as a 4D tensor). The higher order dictionary contains multiple smaller matrices (called sub-dictionaries), so that not only spatial and nonlocal correlation but also spectral correlation can be taken into account.

The sparsity regularization promotes the sparsity of the signal to be recovered with respect to the selected sparsifying basis. In this paper, we assume that we can roughly estimate magnitudes of sparse coefficients, for example, from an intermediately recovered image in iterative recovery. Accordingly, we propose to recover those coefficients whose magnitudes are expected to be large as large values, while recovering the coefficients expected to be small in magnitude as small values. It promotes the sparsity of the recovered sparse representation. This concept is implemented with a weight tensor in a sparsity regularization, called weighted sparsity regularization, where large weight values are assigned to those coefficients expected to have large magnitude and vice versa. We formulate the higher order dictionary learning and spare representation recovery as an optimization problem which can be solved iteratively by the split Bregman method [26]. In this iterative solution, the weighted sparsity regularization is interpreted as a weighting process and, idealistically, the weight tensor should be designed in such a way that the weighting process works as the minimum mean squared error (MMSE)

The authors are with the School of Electronic and Electrical Engineering, Sungkyunkwan University, Korea.

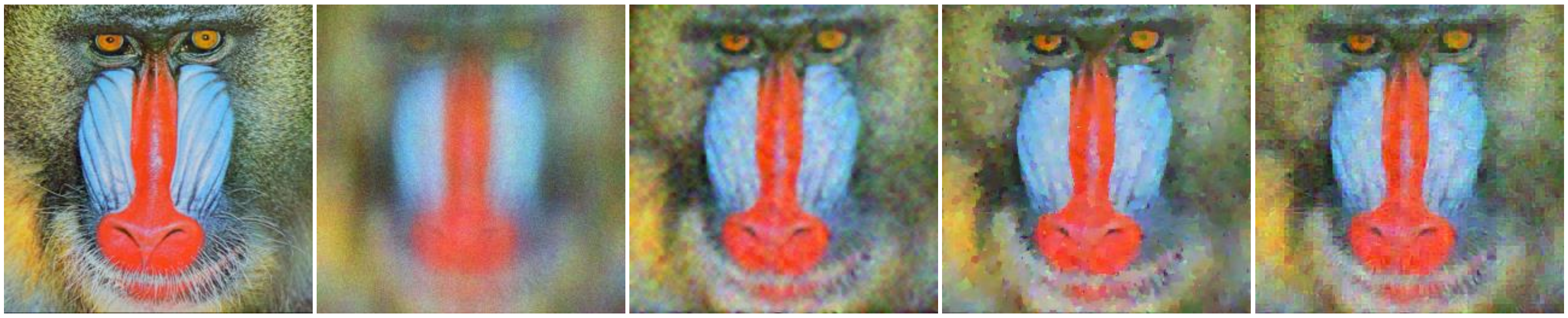

Original Mandrill image | $\ell_1$ [1] (DCT, $\ell_1$-norm) (19.02 dB, 0.647) | SPL [12] (wavelet, $\ell_0$-norm) (21.77 dB, 0.732) | TVAL3 [11] (gradient, $\ell_1$-norm) (21.64 dB, 0.698) | FPD [29] (trained dictionary, $\ell_1$-norm) (21.48 dB, 0.747)

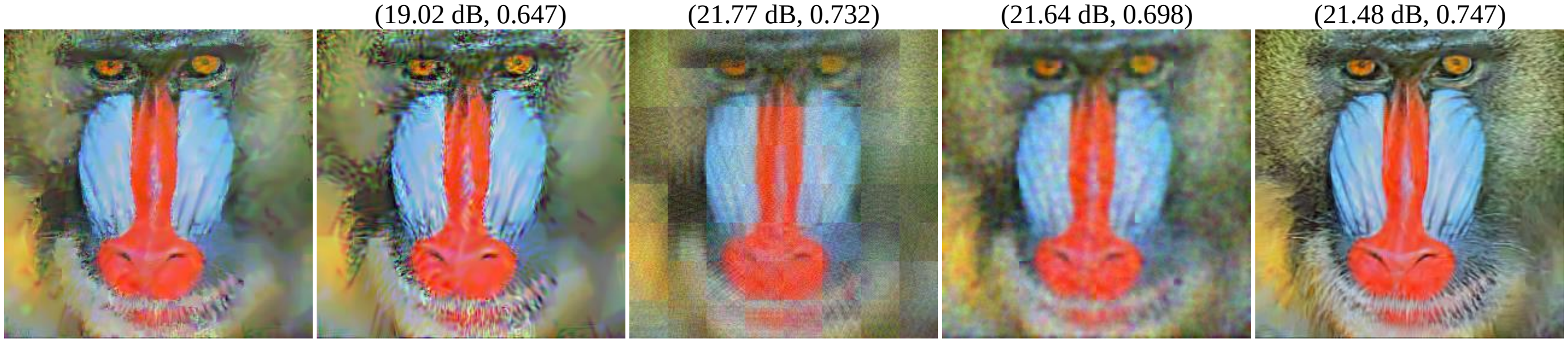

NLR [14] (nonlocal dictionary, log-norm) (21.70 dB, 0.737) | GSR [15] (nonlocal dictionary, $\ell_0$-norm) (20.19 dB, 0.739) | SL20 [17] (DCT, $\ell_{20}$-norm) (19.94 dB, 0.742) | SGIHT [18] (DCT, $\ell_{20}$-norm (21.69 dB, 0.723) | Proposed HODW$_{q=\infty}$ (higher order dictionary, weight-norm) (24.43 dB, 0.867)

Figure 1. Comparison of recovered color images by different recovery methods employing different sparsity bases and sparsity regularizations. Numbers in brackets are PSNR and FSIMc [46], respectively. Sensing matrix is a structural sensing matrix [45] at subrate 0.1.

filter [27] to bring the optimal solution. However, since the true sparse representation is not available at recovery, in practice, we design the weight tensor to approximate the MMSE filter with several variations of soft thresholding, weighted soft thresholding, and hard thresholding.

Contributions of this paper is summarized as followings: 1) to the best of the authors' knowledge, this is the first paper which uses the HOSVD to develop a higher order dictionary for CS problem of color images that 2) adapts to each group of nonlocal similar patches by considering each patch as a high dimensional tensor; 3) and accompanies the weighted sparsity regularization using a weight tensor to faithfully recover the sparse representation generated by the higher order dictionary, in which 4) the weight tensor is designed in several different ways to approximate the MMSE filter.

The rest of this paper is organized as following. Section II discusses prior works, whereas Sections III and IV present formations of CS of color images with the higher order dictionary accompanied by the weighted sparsity regularization, and its iterative solution via split Bregman method, respectively. Section V discusses design of the weight tensor and its variations. Section VI gives experimental results and discussions, while Section VII concludes the paper.

## II. Prior Work

### A. Sparsifying basis and sparsity regularization

CS of images typically assumes that the given image is sparse in fixed sparsifying bases such as 2D-DCT or wavelet [12], [28]. It is also noted that the image gradient can also make the image sparsified, and it is employed in the total variation approach [11] which has been shown to well present the edge and detail information. Xu *et al.* [29] designed a sparsifying basis for natural images, where the sparsifying basis is considered as a dictionary trained from an image dataset to better describe the natural images than the fixed transform bases. However, it should be noted that not only the fixed transform bases or gradient, but also the trained dictionary lack in adaptability to variety of image properties due to the limit of the dataset. On the other hand, the state-of-the-art recovery methods such as low-rank minimization [14] or group-based sparsity [15] try to exploit the nonlocal property of images to learn a specific dictionary adapted for each group of similar patches. Although the works [14], [15] achieved better performance than the fixed sparsifying bases as 2D-DCT, wavelet, gradient, and trained dictionary as shown in Fig. 1 and gave a glimpse of hint on adaptive dictionary learning by exploiting nonlocal properties of images, they focused on gray images by considering each patch as a 1D vector, thus cannot produce the best performance for color images having additional similarity in spectral dimension. We overcome this problem by treating each patch as a 3D tensor so that the learned dictionary can address all the correlation within a group of similar patches including color similarity.

On the other hand, exploitation of prior knowledge on the signal of interest can improve CS recovery. Reweighted $\ell_1$-minimization [30] which assumes certain knowledge on magnitude of the sparse representation sets the weight values in the $\ell_1$ regularization to be inversely proportional to the coefficients' magnitude - it counteracts its influence on the $\ell_1$-norm so that the sparsity of the recovered signal is promoted. This scheme is proven in [30] to be equivalent to regularization with a sparsity regularization using the log-norm. Weighted nonlocal low-rank method [14] also employed the sparsity regularization of log-norm with a sparse representation containing set of singular values of each group of similar patches. Different from the reweighted $\ell_1$-minimization [30] that requires complete recovery to estimate the weight values, the work [14] updates the weight values at every iteration using the recovered data at previous iteration. It can drastically save computation while causing almost no loss in recovery performance. By the way, typical CS recovery methodologies have data fidelity constraint and sparsity regularization, and their optimization process can be considered as alternate iterations of a data update process (to address the data fidelity constraint) and a filtering process (to do regularization on the sparsity) that estimates the signal of interest from output of the update process. Therefore, CS recovery can be improved through removing perturbation in the recovered image at the current iteration by filtering with Wiener filter in smooth Landweber projection [12], Wiener filter in $\ell_{20}$ minimization [31], or some other sophisticated filters (e.g., BM3D filter [32]) in approximate message passing (AMP) [33]. This paper shares the same concept of promoting sparsity of the sparse representation so that large-magnitude coefficients can be recovered as large ones and vice versa as in [30] by a weighted sparsity regularization accompanied by a weight tensor for the case of higher order dictionary. Furthermore, while [14] and [30] designed the weights that counteract the coefficients magnitude to pursue sparsity by log-norm regularization, the weighted sparsity regularization is more general and offers more freedom in designing the weight tensor. We analyze its recovery error at each iteration and then design the weight tensor in several variations to approximate the minimum mean squared error (MMSE) filter.

### B. *Color correlation exploitation*

Due to the ever-increasing importance of vivid color information in modern media systems, more attention is given to CS recovery of color images for which color correlation has been typically taken into account by simultaneous recovery of three color channels using their CS measurements. Nagesh *et al.* [16] recovered color images based on a joint sparsity model that simultaneously processed so called, common and innovation components of red (*R*), green (*G*), and blue (*B*) channels. Majumdar and Ward [17] observed that if *R* value is large, then *G* and *B* values are also large due to strong correlation among three color channels, which led them to use $\ell_{2,0}$ to exploit this property. This $\ell_{2,0}$ framework is improved by an embedded Wiener filter in [31] or combined with the iterative hard thresholding in [18]. However, since these works used fixed sparsifying bases (i.e., 2D-DCT) which cannot adapt to nonlocal property of images, they cannot show satisfactory performance, as seen in Fig. 1. Compared to them, we utilize the nonlocal property of images to learn an adaptive higher order dictionary by considering a group of nonlocal similar patches as a 4D tensor of three color channels so that the higher order dictionary can decompose all correlation within the tensor including the color correlation.

## III. Higher order dictionary and sparsity regularization

### A. *Problem formulation*

Denote $X \in \mathbb{R}^{h\times w\times 3}$ an original color image having spatial resolution of $h \times w$ and three color channels of *R*, *G*, and *B*. CS takes a measurement vector *y* of *X* in a sensing process as:

$$y = \Phi x . \tag{1}$$

Here $x = \left[ x_R^T \quad x_G^T \quad x_B^T \right]^T$ is a vectorized concatenation of three *R*, *G*, *B* color channel vectors $x_R, x_G, x_B$ of the image *X*. $y = \left[ y_R^T \quad y_G^T \quad y_B^T \right]^T$ is the corresponding measurement vector of *x*, in which $y_C = \Phi_C x_C$ , $C$ = *R*,*G*,*B*, and $\Phi_C$ is a sensing matrix for the color channel *C*. Accordingly, $\Phi$ is modeled as $\Phi = diag\left(\Phi_R, \Phi_G, \Phi_B\right)$. The ratio of number of measurements in *y* to the number of pixels in *x* is called the subrate (or measurement rate) which represents sampling rate reduction by CS compared to the conventional Nyquist/Shannon rate.

CS recovery is to faithfully recover the image vector *x* from the ill-posed problem (1). Given a sparsifying basis *D* that sparsifies *x* into a sparse representation $\alpha$, $x = D\alpha$, CS recovers the sparse representation $\alpha$ using an optimization method with a data constraint on the measurement *y* and a regularization on the sparsity of $\alpha$ , $\|\alpha\|_\#$, as:

$$\hat{x} = \arg\min_x \frac{1}{2}\|y - \Phi x\|_2^2 + \lambda \|\alpha\|_\# \quad s.t.\ \ x = D\alpha . \tag{2}$$

Conventionally, the sparsity regularization $\|\alpha\|_\#$ can be $\ell_0$-, $\ell_1$-, or log-norm. (2) shows that performance of the CS recovery depends on how sparse the sparse representation $\alpha$ can be made by the sparsifying basis *D*; and how faithfully $\alpha$ can be recovered. Therefore, CS has to solve two basic problems simultaneously: one is finding a good sparsifying basis *D* that can decorrelate the signal of interest to produce a sparser $\alpha$; and the other is finding a proper sparse regularization $\|\alpha\|_\#$ that can bring a more faithful recovery of $\alpha$ via an optimization method of choice. This paper addresses these issues by using the HOSVD [19] to adaptively learn a dictionary (i.e., sparsifying basis) for the image $x$ and using a weighted sparsity regularization to recover the corresponding sparse representation $\alpha$ .

### B. *Nonlocal higher-order dictionary*

State-of-the-art image restoration schemes [14], [15], [34], [35] improve restoration performance using similarity among a group of similar patches within an image, which is based on spatial similarity among values of pixels in a patch, and nonlocal similarity among values of pixels in different patches spatially not necessarily adjacent each other. However, another similarity exists for color images, that is, the similarity among color channels (e.g., *R*, *G* and *B* values), namely spectral similarity [36], [37]. We show this similarity in terms of cross correlation among any two color channels as in Fig. 2, where the cross correlation is calculated as, for example of *R* and *G* channel, $\rho_{RG} = \frac{(x_R - m_R)^T (x_G - m_G)}{\|x_R - m_R\|_2 \|x_G - m_G\|_2}$ with $m_R$, $m_G$ denoting respectively average values of $R$ and $G$ channels. The correlation between any two color channels is high, which is usually higher than 0.8. Therefore, to exploit all of these similarities including spectral similarity for CS of color images, we propose to use the HOSVD [21], [24], [25], [38] to decompose these similarities so that we can learn a higher order dictionary adaptively for each group of color patches for special case of compressive sensing of color images. We then set up an algorithm to recover a color image from CS measurements using a proper regularization on the sparsity of the sparse representation corresponding to the learned dictionary to improve quality of the recovered color image.

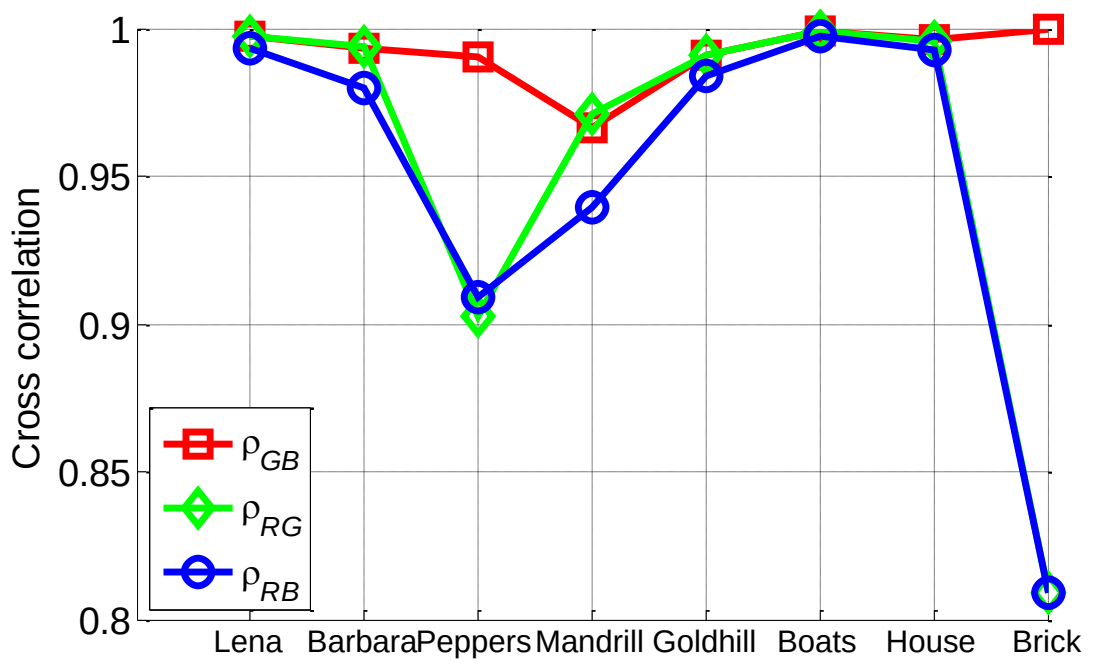

Figure 2. Cross correlation among color channels for various test images.

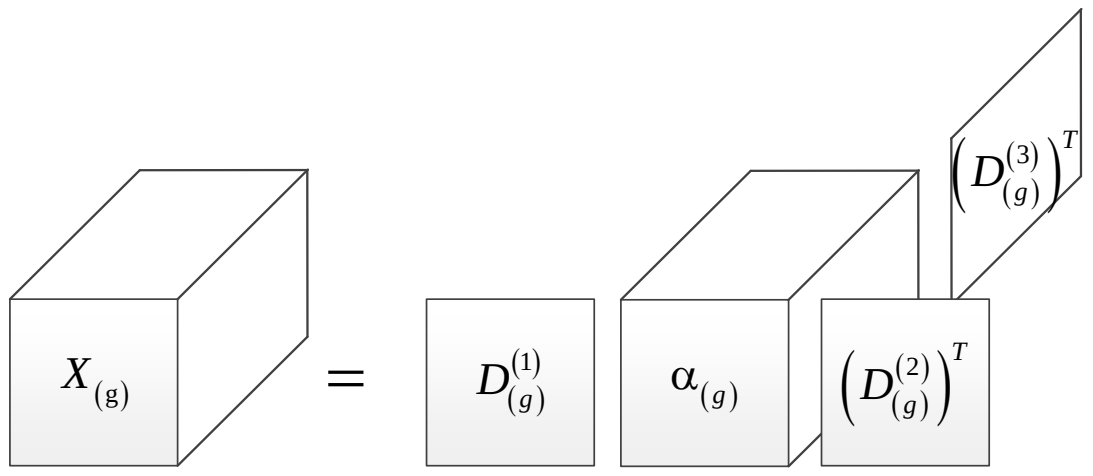

Figure 3. Visualization of a dictionary $\alpha_{(g)}$, $k=1,\ldots,K$ generated from tensor $X_{(g)}$ using HOSVD in case of $K$ = 3, but it can be generalized to a larger $K$.

For the image $x$, we extend the notations used in literature [21], [24], [25], [38] to denote $R_g(x), g=1,..,N_g$, the operator that extracts the $g$-th group of $L$ similar patches of size $p \times p \times 3$, $X_{(g)} \in \mathbb{R}^{p \times p \times 3 \times L}$, from the image $x$; and $R_g^T\left(X_{(g)}\right)$ is the corresponding inverse operation that aggregates the group $X_{(g)}$ back to its corresponding positions in the vector $x$ and pads zeros elsewhere. $N_g$ is the number of groups in the image $x$. We then apply the HOSVD to each $X_{(g)}$ to find the higher order singular vectors, denoted by $D_{(g)}$, which is then considered as a dictionary for $X_{(g)}$. Because $X_{(g)}$ has a dimension of $K$ = 4, $D_{(g)}$ composes of $K$ = 4 matrices, considered as sub-dictionaries, that is, $D_{(g)} = \left\{D^{(k)}_{(g)}\right\}, k=1,..,K$. Each sub-dictionaries $D^{(k)}_{(g)}$ decomposes the similarity of the group $X_{(g)}$ in the corresponding ($k$-th) dimension. We illustrate this concept in Fig. 3, however, for the case $K$ = 3. According to the dictionary $D_{(g)}$, the sparse representation is (higher order) singular values $\alpha_{(g)} \in \mathbb{R}^{p \times p \times 3 \times L}$ such that $X_{(g)} = \alpha_{(g)} \times_{k=1}^{K} D^{(k)}_{(g)}$, where $\times_{k=1}^{K}$ is an operator that reconstructs the group $X_{(g)}$ from its higher order singular values $\alpha_{(g)}$ and higher order singular vectors $D_{(g)} = \left\{D^{(k)}_{(g)}\right\}$, $k=1,..,K$. Therefore, for the whole image $x$, we can form a higher order dictionary as $D = \left\{D_{(g)}\right\}, g=1,..,N_g$ with its corresponding spare representation as $\alpha = \left\{\alpha_{(g)}\right\}$, $g=1,..,N_g$. The whole image $x$ can be retrieved as $x = \alpha \bullet D$ as following:

$$x = \alpha \bullet D \triangleq \sum_{g=1}^{N_g} R_g^T\left(\alpha_{(g)} \times_{k=1}^{K} D^{(k)}_{(g)}\right) ./ \sum_{g=0}^{N_g} R_g^T\left(1_{p\times p\times 3\times L}\right), \quad (3)$$

where ./ is the element-wise division and $1_{p\times p\times 3\times L}$ is a tensor with all values of 1's. The operation of $x = \alpha \bullet D$ in (3) retrieves $N_g$ groups of patches $X_{(g)} = \alpha_{(g)} \times_{k=1}^{K} D^{(k)}_{(g)}$ from the corresponding sparse representation $\alpha_{(g)}$ and aggregates all groups of $X_{(g)}$ to the vector $x$.

Because the image $x$ is not available at the recovery, instead of learning the dictionary $D$ from $x$, we learn it from $x^{ref}$ denoting its reference (which can be erroneous) version of $x$. Availability of $x^{ref}$ will be discussed latter.

### C. *Weighted sparsity regularization*

To recover the sparse representation $\alpha = \left\{\alpha_{(g)}\right\}, g=1,..,N_g$, we need regularization on its property (i.e., sparsity). We assume that side information on the magnitudes of coefficients of the sparse representation is available at recovery. For example, this information can be estimated from a sparse representation $\alpha^{ref}$ (regarding the dictionary $D$) generated from the reference $x^{ref}$, as $x^{ref} = \alpha^{ref} \bullet D$. Accordingly to side information, we can recover those coefficients expected to be large in magnitude as large ones in magnitude and vice versa, by defining a weighted sparsity regularization that uses a high dimensional weight tensor, $W_{(g)} \in \mathbb{R}^{p\times p\times 3\times L}$ as:

$$\left\|\alpha_{(g)}\right\|_W \triangleq \left\|W_{(g)} \circ \alpha_{(g)}\right\|_F^2 = \sum_{i_1} \ldots \sum_{i_K} \left(W_{(g)|i_1,\ldots,i_K} \alpha_{(g)|i_1,\ldots,i_K}\right)^2, \quad (4)$$

where $W_{(g)}$ has non-negative weight values; $\|s\|_F$ is an extension of the Frobenious norm [39] to calculate square root of summation of all squared elements of a tensor $s$; the operator $\circ$ denotes the Hadamard product (i.e., element-wise multiplication); and $i_1,\ldots,i_K$ are indices of elements of tensors $W_{(g)}$ and $\alpha_{(g)}$. Subsequently, sparsity regularization of the whole image is summation of all groups:

$$\|\alpha\|_W = \sum_{g=1}^{N_g} \left\|\alpha_{(g)}\right\|_W . \quad (5)$$

In this definition, we assign a large weight value $W_{(g)|i_1,\ldots,i_K}$ to the coefficient $\alpha_{(g)|i_1,\ldots,i_K}$ which is expected to have small magnitude so that it is minimized better by the minimization in (2) to be recovered as a small one in magnitude. On the other hand, a small weight value $W_{(g)|i_1,\ldots,i_K}$ is given to the coefficient $\alpha_{(g)|i_1,\ldots,i_K}$ whose magnitude is expected to be large for a possibly large-magnitude recovered value.

## IV. Iterative Solution via Split Bregman Method

The optimization problem in (2) is rewritten in terms of the higher order dictionary and the weighted sparsity regularization as:

$$\hat{x} = \arg\min_x \frac{1}{2}\|y - \Phi x\|_2^2 + \lambda\|\alpha\|_W \quad s.t. \quad x = \alpha \bullet D. \tag{6}$$

By using the split-Bregman method [26], which introduces an auxiliary variable *b*, (6) can be iteratively solved as following:

$$x^{(t+1)} = \arg\min_x \frac{1}{2}\|y - \Phi x\|_2^2 + \frac{\mu}{2}\left\|x - \alpha^{(t)} \bullet D - b^{(t)}\right\|_2^2, \tag{7}$$

$$\alpha^{(t+1)} = \arg\min_\alpha \lambda\|\alpha\|_W + \frac{\mu}{2}\left\|x^{(t+1)} - \alpha \bullet D - b^{(t)}\right\|_2^2, \tag{8}$$

$$b^{(t+1)} = b^{(t)} - \left(x^{(t+1)} - \alpha^{(t+1)} \bullet D\right). \tag{9}$$

### A. *x subproblem in (7)*

By setting derivative of the cost function of (7) to zero, the solution of *x* is:

$$x^{(t+1)} = \left(\Phi^T\Phi + \mu I\right)^{-1}\left(\Phi^T y + \mu\left(\alpha^{(t)} \bullet D + b^{(t)}\right)\right). \tag{10}$$

The large size of the sensing matrix $\Phi$ makes direct calculation of the solution of (10) hard. However, it can be solved efficiently by approximating the inverse matrix $\left(\Phi^T\Phi + \mu I\right)^{-1}$ with the Neumann series [39], or by solving the subproblem in (7) using the gradient decent method [40]. This paper uses the gradient decent method, with initial $x^{(t+1)} = 0$, to solve $x^{(t+1)}$ iteratively (until convergence) as:

$$x^{(t+1)} = x^{(t+1)} - \eta d, \tag{11}$$

with the step size $\eta$ and the gradient direction *d* of:

$$d = \left(\Phi^T\left(\Phi x^{(t+1)} - y\right) + \mu\left(x^{(t+1)} - \alpha^{(t)} \bullet D - b^{(t)}\right)\right). \tag{12}$$

### B. *α subproblem in (8)*

#### 1) *Subproblem decomposition*

The subproblem in (8) contains $\left\|x^{(t+1)} - \alpha \bullet D - b^{(t)}\right\|_2^2$, thus it is not easy to find its derivative. Therefore, we recast the subproblem, by extending the Theorem 1 in [15], as a case of the higher order dictionary for an easy solution. We denote $r^{(t+1)}$, the sparser representation of the (already available) residual $x^{(t+1)} - b^{(t)}$ regarding the dictionary *D* as:

$$x^{(t+1)} - b^{(t)} = r^{(t+1)} \bullet D. \tag{13}$$

We assume the elements of the residual $e = x^{(t+1)} - \alpha \bullet D - b^{(t)}$ are independent of each other and have zero mean and variance $\sigma_e^2$. Therefore, by the law of large number [41], for any

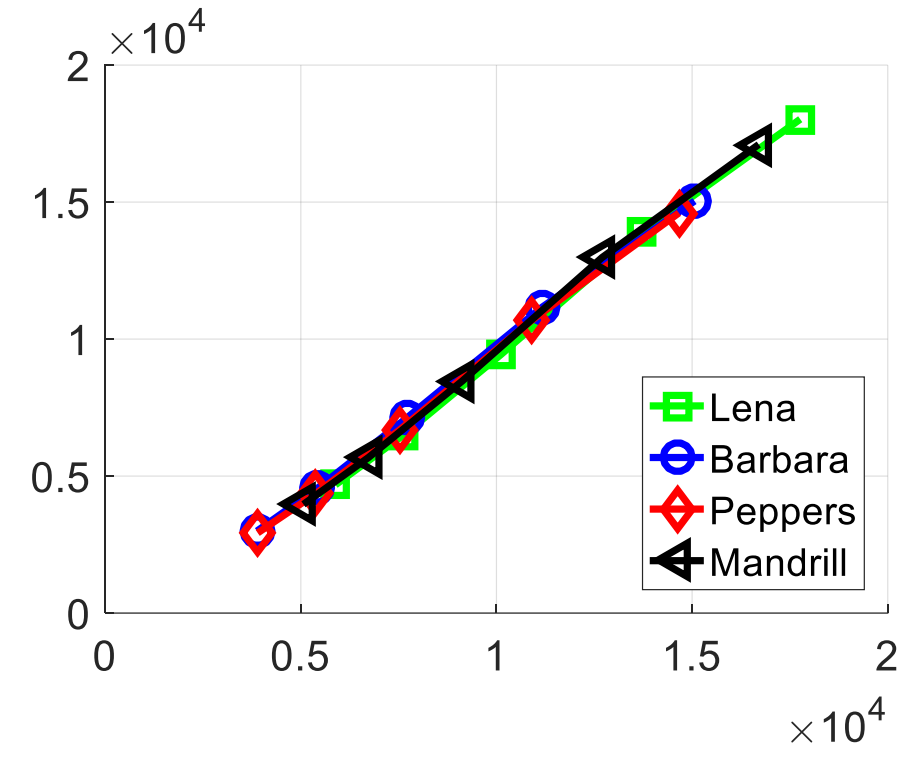

Figure 4. Comparison of $\frac{1}{3hw}\left\|x^{(t+1)} - \alpha \bullet D - b^{(t)}\right\|_2^2$ (in vertical axis) and $\frac{1}{3N_g p^2 L}\sum_{g=1}^{N_g}\left\|r_{(g)}^{(t+1)} - \alpha_{(g)}\right\|_2^2$ (in horizontal axis) for different iteration *t* = 1, 10, 20, 30, 40, 50 (respectively from top-right to left-bottom markers) at subrate 0.1, where $\alpha$ is taken from the ground-truth *x* for several test images.

arbitrarily small $\varepsilon$, the following inequality holds in probability:

$$\underset{3hw\to\infty}{Lim}\, P\left\{\left|\frac{1}{3hw}\left\|x^{(t+1)} - \alpha \bullet D - b^{(t)}\right\|_2^2 - \sigma_e^2\right| < \frac{\varepsilon}{2}\right\} = 1, \tag{14}$$

where the total number of elements of error $e = x^{(t+1)} - \alpha \bullet D - b^{(t)}$ is $3hw$. By further assuming that elements of the *g*-th group $E_{(g)}, g = 1,..,N_g$, extracted from *e* are independent of each other, elements of its sparse representation (i.e., singular values, $r_{(g)}^{(t+1)} - \alpha_{(g)}, g = 1,..,N_g$) also have zero mean and variance $\sigma_e^2$. It is because the dictionary $D_{(g)}$ has its sub-dictionaries of unitary matrices (as they are singular vectors of *g*-th group of $x^{ref}$). Also by the law of large number, we can conclude:

$$\underset{3N_g p^2 L\to\infty}{Lim}\, P\left\{\left|\frac{1}{3N_g p^2 L}\sum_{g=1}^{N_g}\left\|r_{(g)}^{(t+1)} - \alpha_{(g)}\right\|_2^2 - \sigma_e^2\right| < \frac{\varepsilon}{2}\right\} = 1, \tag{15}$$

where the total number of elements of $r_{(g)}^{(t+1)} - \alpha_{(g)}, g = 1,..,N_g$, is $3N_g p^2 L$. From (14) and (15) we have the approximation with high probability when the size of the image is sufficiently large:

$$\frac{1}{3hw}\left\|x^{(t+1)} - \alpha \bullet D - b^{(t)}\right\|_2^2 \approx \frac{1}{3N_g p^2 L}\sum_{g=1}^{N_g}\left\|r_{(g)}^{(t+1)} - \alpha_{(g)}\right\|_2^2. \tag{16}$$

Figure 4 confirms the approximation in (16) for the desired case of $\alpha$ taken from the original image vector *x*, for different test images at different iterations. Accordingly, the subproblem (8) can be approximated by the following decomposition:

$$\alpha^{(t+1)} = \arg\min_\alpha \sum_{g=1}^{N_G}\left\|\alpha_{(g)}\right\|_W + \mu'\sum_{g=1}^{N_g}\left\|r_{(g)}^{(t+1)} - \alpha_{(g)}\right\|_2^2, \tag{17}$$

where all constants are gathered within $\mu'$ as:

$$\mu' = \mu hw / 2\lambda L N_g p^2. \tag{18}$$

Accordingly to the additive property of the norm $\|\alpha\|_W$ in (5), the problem in (17) can be efficiently decomposed into independently solvable $N_g$ problems:

$$\alpha_{(g)}^{(t+1)} = \min_{\alpha_{(g)}} \mu' \left\| r_{(g)}^{(t+1)} - \alpha_{(g)} \right\|_2^2 + \left\| \alpha_{(g)} \right\|_W . \quad (19)$$

*2) Optimization for a group of patches*

By definition in (4), sparsity regularization $\left\|\alpha_{(g)}\right\|_W$ has its derivative to an $(i_1,...,i_K)$ element as:

$$\frac{\partial \left\|\alpha_{(g)}\right\|_W}{\partial \alpha_{(g)|i_1,...,i_K}} = \frac{\partial \sum_{i_1}...\sum_{i_K}\left(W_{(g)|i_1,...,i_K}\alpha_{(g)|i_1,...,i_K}\right)^2}{\partial \alpha_{(g)|i_1,...,i_K}} = 2\left(W_{(g)|i_1,...,i_K}\right)^2 \alpha_{(g)|i_1,...,i_K} . \quad (20)$$

By computing derivative to all elements $(i_1,...,i_K)$ in the same manner in (20) and rearranging its results to the shape of $\alpha_{(g)}$, we have the derivative of $\left\|\alpha_{(g)}\right\|_W$ as,

$$\frac{\partial \left\|\alpha_{(g)}\right\|_W}{\partial \alpha_{(g)}} = 2W_{(g)} \circ W_{(g)} \circ \alpha_{(g)} . \quad (21)$$

Similarly, derivative of $\left\| r_{(g)}^{(t+1)} - \alpha_{(g)} \right\|_F^2$ respect to $\alpha_{(g)}$ is:

$$\frac{\partial \left\| r_{(g)}^{(t+1)} - \alpha_{(g)} \right\|_F^2}{\partial \alpha_{(g)}} = 2\left(\alpha_{(g)} - r_{(g)}^{(t+1)}\right). \quad (22)$$

Accordingly, by setting the derivative of cost function in (19) regarding $\alpha_{(g)}$ to zero, it has a solution of:

$$\begin{aligned} &\mu' 2\left(\alpha_{(g)} - r_{(g)}^{(t+1)}\right) + 2W_{(g)} \circ W_{(g)} \circ \alpha_{(g)} = 0 \\ &\Leftrightarrow \left(W_{(g)} \circ W_{(g)} + \mu'\right) \circ \alpha_{(g)} = r_{(g)}^{(t+1)}. \end{aligned} \quad (23)$$

(23) shows that an $(i_1,...,i_K)$ coefficient of $\alpha_{(g)}^{(t+1)}$ in (19) is calculated by the following equation:

$$\alpha_{(g)|i_1,...,i_K}^{(t+1)} = \frac{r_{(g)|i_1,...,i_K}^{(t+1)}}{W_{(g)|i_1,...,i_K}^2 + \mu'} . \quad (24)$$

The coefficients of $\alpha^{(t+1)}$ are weighted by weights calculated using the corresponding coefficients of $r^{(t+1)}$.

## V. Design of Weight Tensor for Sparsity Regularization

### A. Design of weight tensor

*Interpretation of the weighting process*. Because the concept of recovery in (6) is to find a solution under the data constraint and the sparsity regularization, we consider the solution of (6) as iterations of two basic processes of update and filtering. The update process is modeled by the *x* subproblem in (7), in which the current recovered image is updated using information from the measurements as in (10). The filtering process as modeled in the *α* subproblem in (8) is to promote the recovered sparse representation to have the image property defined in the sparsity regularization, or in other words, to filter the sparse representation to its true one by the weighting process in (24). Therefore, the weight tensor *W* can be designed from the filtering perspective, by which the weighting process in (24) can bring a better approximation to the true sparse representation.

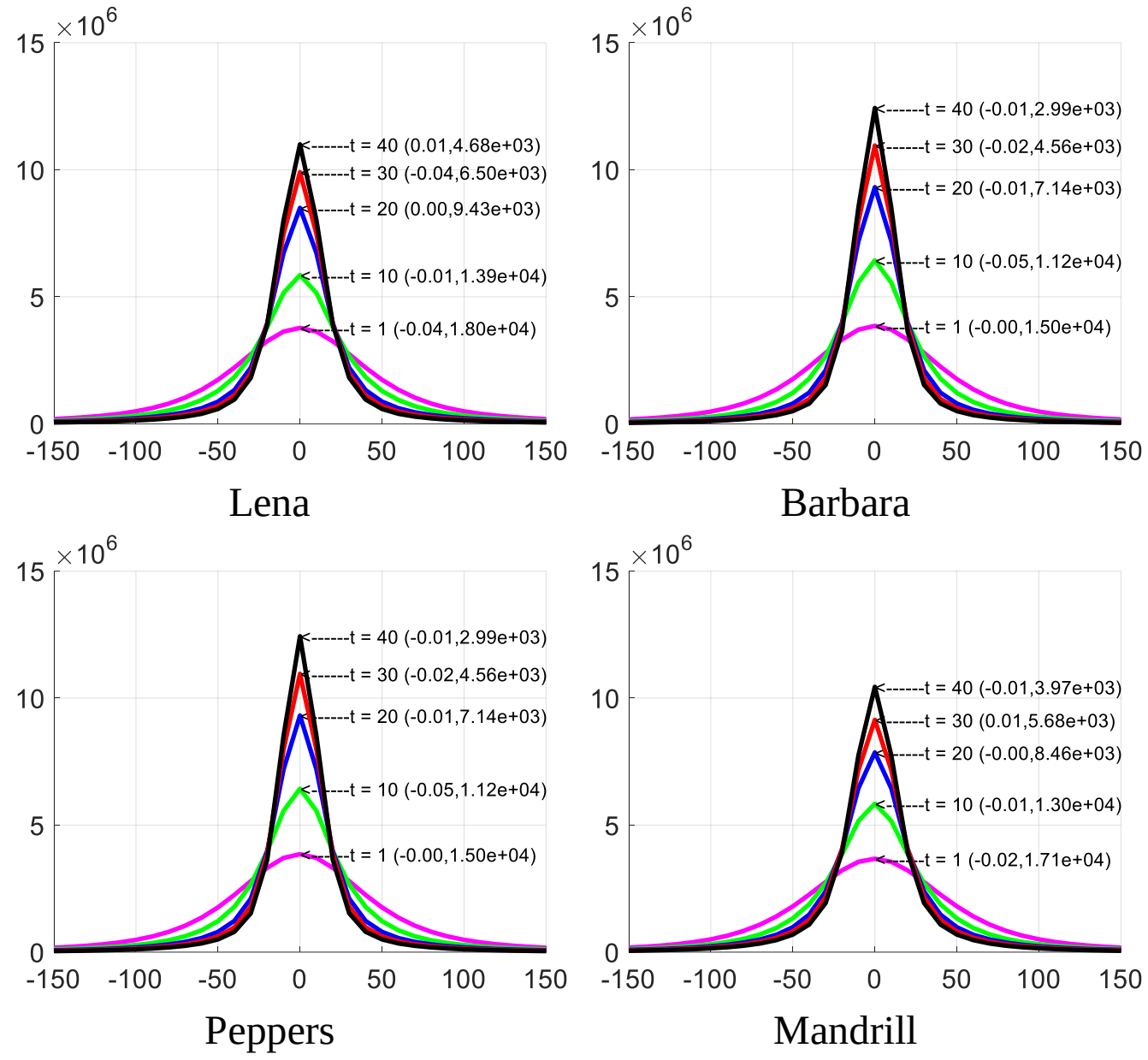

Figure 5. Histogram of error $r_{(g)}^{(t+1)} - \alpha_{(g)}^0, g = 1,..,N_g$ by the proposed HODW at different iteration *t*, subrate *r* = 0.1. Numbers in bracket are mean and variance values of the error, respectively.

*Sparse representation error*. Suppose the true group $X_{(g)}$ has the true sparse representation $\alpha_{(g)}^0$ regarding the dictionary $D_{(g)}$, that is, $X_{(g)} = \alpha_{(g)}^0 \times_{k=1}^K D_{(g)}^{(k)}$. $r_{(g)}^{(t+1)}$ is then considered as an erroneous observation of $\alpha_{(g)}^0$ with error of $r_{(g)}^{(t+1)} - \alpha_{(g)}^0$ at the *t*-th iteration. To filter the sparse representation $r_{(g)}^{(t+1)}, g = 0,..., N_g$, toward the true $\alpha_{(g)}^0$, we assume its error of $\left(r_{(g)}^{(t+1)} - \alpha_{(g)}^0\right)$ has zero mean and standard deviation $\sigma^{(t+1)}$. This assumption is verified in Fig. 5 which confirms that the sparse representation error has very small mean value (i.e., close to zero) irrespective of images at different iterations.

*Oracle design of weight tensor*. From the filtering perspective, the weight values in the weighting process in (24) should be designed by minimum mean squared error (MMSE) filter perspective as:

$$\alpha^{(t+1)} = \arg\min_{\alpha} E\left[\left(\alpha - \alpha^0\right)^2\right], \quad (25)$$

where $E[s]$ is the expected value taken over all elements $i_1,...,i_K$ of all group $g = 1,..,N_g$ of *s*. Similar to Wiener filter [27], however, applied to higher order sparse representation, the optimal solution for an element of $\alpha_{(g)|i_1,...,i_K}^{(t+1)}$ by (25) regarding the standard deviation of error $\sigma^{(t+1)}$ is:

TABLE I
THE PROPOSED HODW FOR COMPRESSIVE SENSING OF COLOR IMAGES.

**Input:** measurement $y$, $\Phi$, $\Phi^T$, $\hat{x}_{out}^{(0)}$, $\sigma^*$, $p$, $L$, $N_g$
**for** $tt$ = 1: Outloop
  % *Learn higher order dictionary and weight tensor*
  Extract $N_g$ groups from $x^{ref} = \hat{x}_{out}^{(tt)}$
  Update dictionary $D$ and sparse representation $\alpha^{ref}$ by calculating HOSVD for $N_g$ groups
  Find weight tensor $W$ by eq.(28) from $\alpha^{ref}$
  **for** $t$ = 1:1:Inloop
    % *Update process*
    Update $\hat{x}^{(t)}$ by eq.(10)
    % *Filtering process*
    Filter to find $\alpha^{(t)}$ by eq.(24)
    Update $b^{(t)}$ by eq.(9)
  $\hat{x}_{out}^{(tt)} = x^{(t)}$
**Output:** Clean image $\hat{x}_{out}^{(tt)}$

$$\alpha_{(g)|i_1,\ldots,i_K}^{(t+1)} = \frac{\left(\alpha_{(g)|i_1,\ldots,i_K}^{0}\right)^2}{\left(\alpha_{(g)|i_1,\ldots,i_K}^{0}\right)^2 + \left(\sigma^{(t+1)}\right)^2} r_{(g)|i_1,\ldots,i_K}^{(t+1)} . \quad (26)$$

Accordingly, (24) and (26) infer the weight value of each coefficient of $\alpha_{(g)}^{0}$ as:

$$W_{(g)|i_1,\ldots,i_K} = \begin{cases} \sqrt{\dfrac{\left(\alpha_{(g)|i_1,\ldots,i_K}^{0}\right)^2 + \left(\sigma^{(t+1)}\right)^2}{\left(\alpha_{(g)|i_1,\ldots,i_K}^{0}\right)^2} - \mu'} & if\ \alpha_{(g)|i_1,\ldots,i_K}^{0} \neq 0 \\ \infty & else. \end{cases} \quad (27)$$

We select $\lambda$ such that $\lambda > \mu hw / 2LN_g p^2$, hence, $\mu' < 1$ and a weight value $W_{(g)|i_1,\ldots,i_K}$ exists for any value of $\alpha_{(g)|i_1,\ldots,i_K}^{0}$. This weight design in (27) assigns a small weight $W_{(g)|i_1,\ldots,i_K}$ to large $\left|\alpha_{(g)|i_1,\ldots,i_K}^{0}\right|$, so that it is weighted by a value very close to 1 to be recovered as a large magnitude one as in (26). Reversely, small $\left|\alpha_{(g)|i_1,\ldots,i_K}^{0}\right|$ will be assigned a very large $W_{(g)|i_1,\ldots,i_K}$ (even can go to infinity), and then will be weighted by a very small (close to zero) weight as in (26) to become small.

*Practical design of the weight tensor*. The problem is that the true value of $\alpha_{(g)|i_1,\ldots,i_K}^{0}$ is not available at the recovery. However, with assumption of the availability of the erroneous version $x^{ref}$, we approximate the optimal weight in (27) as:

$$\frac{1}{W_{(g)|i_1,\ldots,i_K}^2 + \mu'} = \begin{cases} \dfrac{\left(\left|\alpha_{(g)|i_1,\ldots,i_K}^{ref}\right|^q - \sigma_*^q\right)_+}{\left|\alpha_{(g)|i_1,\ldots,i_K}^{ref}\right|^q} & if\ \alpha_{(g)|i_1,\ldots,i_K}^{ref} \neq 0 \\ 0 & else. \end{cases} \quad (28)$$

where $\alpha^{ref}$ is a sparse representation of $x^{ref}$ regarding the dictionary $D$, $(s)_+ = \max(s, 0)$, $\sigma_*$ is a positive tuning parameter representing noise level in the sparse representation, and $q$ is a non-negative parameter to control the regularization. Therefore, this is equivalent to a weight tensor designed as:

$$W_{(g)|i_1,\ldots,i_K} = \begin{cases} \sqrt{\dfrac{\left|\alpha_{(g)|i_1,\ldots,i_K}^{ref}\right|^q}{\left|\alpha_{(g)|i_1,\ldots,i_K}^{ref}\right|^q - \sigma_*^q} - \mu'} & if\ \left|\alpha_{(g)|i_1,\ldots,i_K}^{ref}\right|^q > \sigma_*^q \\ \infty & else. \end{cases} \quad (29)$$

It should be noted that the weight tensor $W$ does not need to be explicitly calculated as in (29), but its implicit form in (28) is enough to do the filtering in (24).

### B. Erroneous $x^{ref}$ for learning of D and design of W

Similar to the reweighted $\ell_1$ minimization [30], $x^{ref}$ can be an output of a complete recovery, as shown in Table I. However, it is very expensive to pursue a complete recovery in order to learn the dictionary $D$ and the weight tensor $W$. Therefore, to significantly save the computation of the proposed method, we update the weighted tensor (and then the filtering process) at every iteration, because the improved recovered image at each iteration is expected to lead more accurate learning of the dictionary $D$ and the weight tensor $W$. Furthermore, from the perspective of filtering the spare representation, we can design the filtering process (i.e., weighting process) directly from the signal to be filtered, that is, $r^{(t+1)}$ in (24). Therefore, at each iteration, this paper sets $x^{ref} = x^{(t+1)} - b^{(t)}$ and $\alpha^{ref} = r^{(t+1)}$. In this design, we consider three cases of $q$ (i.e., $q$ = 1, 2, and $\infty$), and its interpretation is given as following.

- With $q$ = 1, by simple derivation, the weight tensor in (28) gives the equivalent filtering of the sparse representation as:

$$\alpha_{(g)|i_1,\ldots,i_K} = sign\left(r_{(g)|i_1,\ldots,i_K}^{(t+1)}\right)\left(\left|r_{(g)|i_1,\ldots,i_K}^{(t+1)}\right| - \sigma_*\right)_+ . \quad (30)$$

This filtering has a similar form to soft thresholding of the sparse representation as low rank minimization [42], [43], where the filtering is performed by sifting all coefficients toward 0.

- With $q$ = 2, another equivalent filtering of the sparse representation is given by the weight tensor in (28):

$$\alpha_{(g)|i_1,\ldots,i_K} = r_{(g)|i_1,\ldots,i_K}^{(t+1)} \frac{\left(\left(r_{(g)|i_1,\ldots,i_K}^{(t+1)}\right)^2 - \sigma_*^2\right)_+}{\left(r_{(g)|i_1,\ldots,i_K}^{(t+1)}\right)^2} . \quad (31)$$

This is basically similar to performing the Wiener filter with the assumption of independent noise of zero-mean and variance $\sigma_*^2$. The filtering can be further interpreted as:

$$\alpha_{(g)|i_1,\ldots,i_K} = sgn\left(r_{(g)|i_1,\ldots,i_K}^{(t+1)}\right)\left(\left|r_{(g)|i_1,\ldots,i_K}^{(t+1)}\right| - \frac{\sigma_*^2}{\left|r_{(g)|i_1,\ldots,i_K}^{(t+1)}\right|}\right)_+ . \quad (32)$$

This has a similar form to the weighted nuclear norm [44], reweighted $\ell_1$ in [30], and weighted low rank [14], all of which perform filtering by sifting all coefficients toward zero by different amount which is inversely proportional to its magnitude.

- Finally, with $q = \infty$, the weight tensor in (28) gives the equivalent filtering of sparse representation as:

$$\alpha_{(g)|i_1,\ldots,i_K} = \begin{cases} r_{(g)|i_1,\ldots,i_K}^{(t+1)} & if\ \left|r_{(g)|i_1,\ldots,i_K}^{(t+1)}\right| > \sigma_* \\ 0 & otherwise. \end{cases} \quad (33)$$

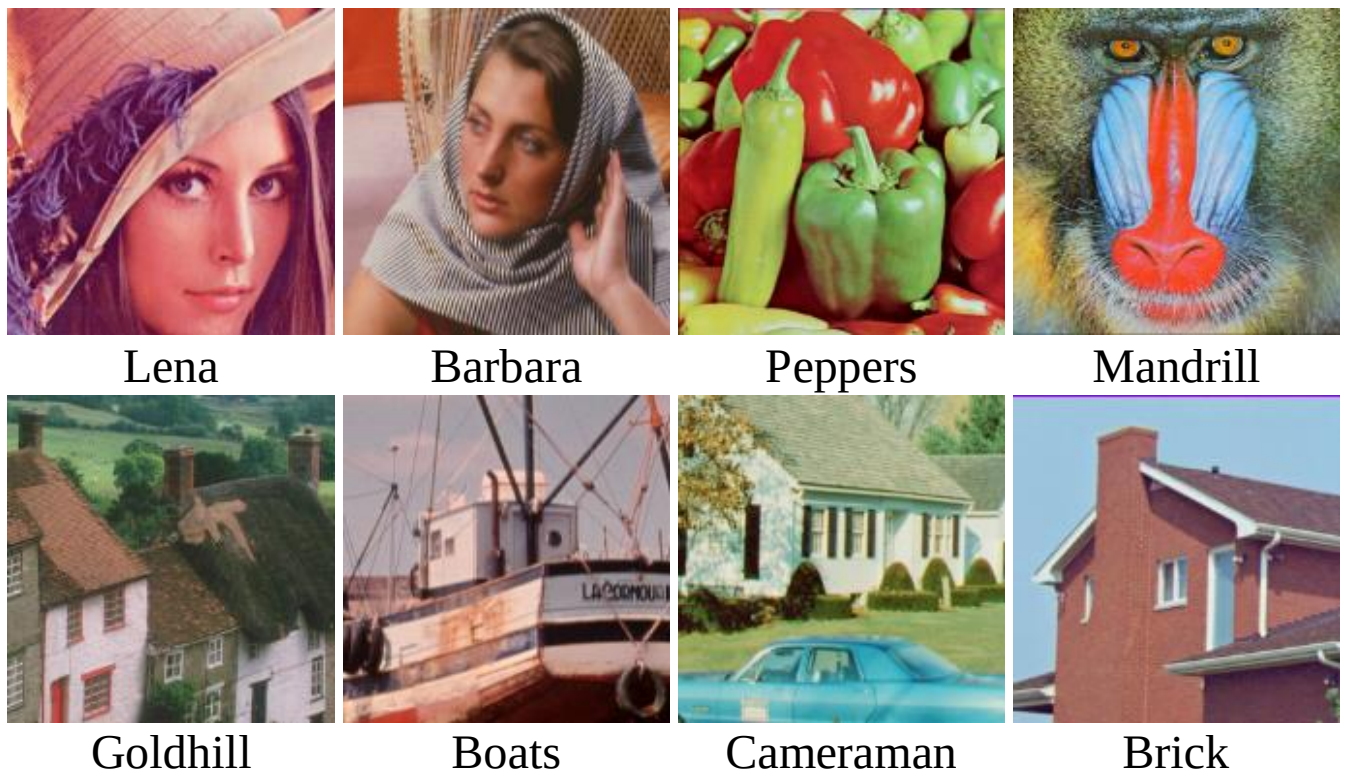

Figure 6. Eight test color images of size 256×256

This is similar to the hard thresholding in regularization with $\ell_0$ norm as in [15], [24]. Therefore, the proposed framework can work in different ways of designing the weight tensor to approximate the MMSE filter. Because of the approximation in (16), the solution of the sub-subproblem in (19) is just an approximation to solution of subproblem in (8). Furthermore, the dictionary $D$ and the weight tensor $W$ are updated at each iteration. Accordingly, the convergence of the proposed method is hard to prove, but we will experimentally show the evolution of the recovery error over iterations in following Section.

## VI. Experimental Results

We test the proposed HODW using 8 color images in Fig. 6 with employing a structural sensing matrix [45] based on Wash-Hadamard matrix for efficient and effective sensing. Four subrates 0.1, 0.2, 0.3, and 0.4 are tested. The parameters of the proposed HODW are set as follows: $\mu = 0.0025$, patch size $p \times p \times c = 8 \times 8 \times 3$, number of patches for each group $L = 60$, which are searched from a $41 \times 41$ search window centered at the current patch. To reduce computation, the current patch is shifted by a distance of 4 pixels both horizontally and vertically to its neighbors. The number of iterations is fixed as $T = 60$. The proposed HODW is tested with three cases of $q$ = 1, 2, ∞, called $\text{HODW}_{q=1}$, $\text{HODW}_{q=2}$, and $\text{HODW}_{q=\infty}$, respectively. Furthermore, we discussed in Section III that we should optimally design the weight tensor $W$ as in (27). Although the true sparse representation is not available at the recovery, we suppose that the optimal weight tensor in (27) is given to the proposed HODW by an oracle, called $\text{HODW}_{\text{Oracle}}$, just to see how much we can further improve the proposed HODW by estimating the weight tensor $W$ very precisely. As a quality assessment index, not only conventional PSNR but also FSIMc [46] (which is shown better than PSNR in imitating human assessment on quality of color images) are reported.

We compare the proposed HODW to existing works accompanying various sets of sparsifying bases and sparsity regularizations such as $\ell_1$-minimization (DCT, $\ell_1$ norm) [1], SPL (DWT, $\ell_0$ norm) [12], TVAL3 (gradient, $\ell_1$ norm) [11], FPD (trained dictionary, $\ell_1$ norm) [29], GSR (nonlocal dictionary, $\ell_0$ norm) [15], and NLR (nonlocal dictionary, *log* norm) [14]. Since they are developed for recovery of gray images, we applied them independently to each channel. We also compared the proposed HODW to other recovery schemes of color images, that is, GSL20 [17] and SGIHT [18]. Noting that the proposed HODW utilizes the nonlocal property of images, its computation can be saved by a warm start (in this paper, it is initiated with the result by TVAL3 [11]). The performance of the proposed HODW without the initial states is investigated later in the performance evaluation section.

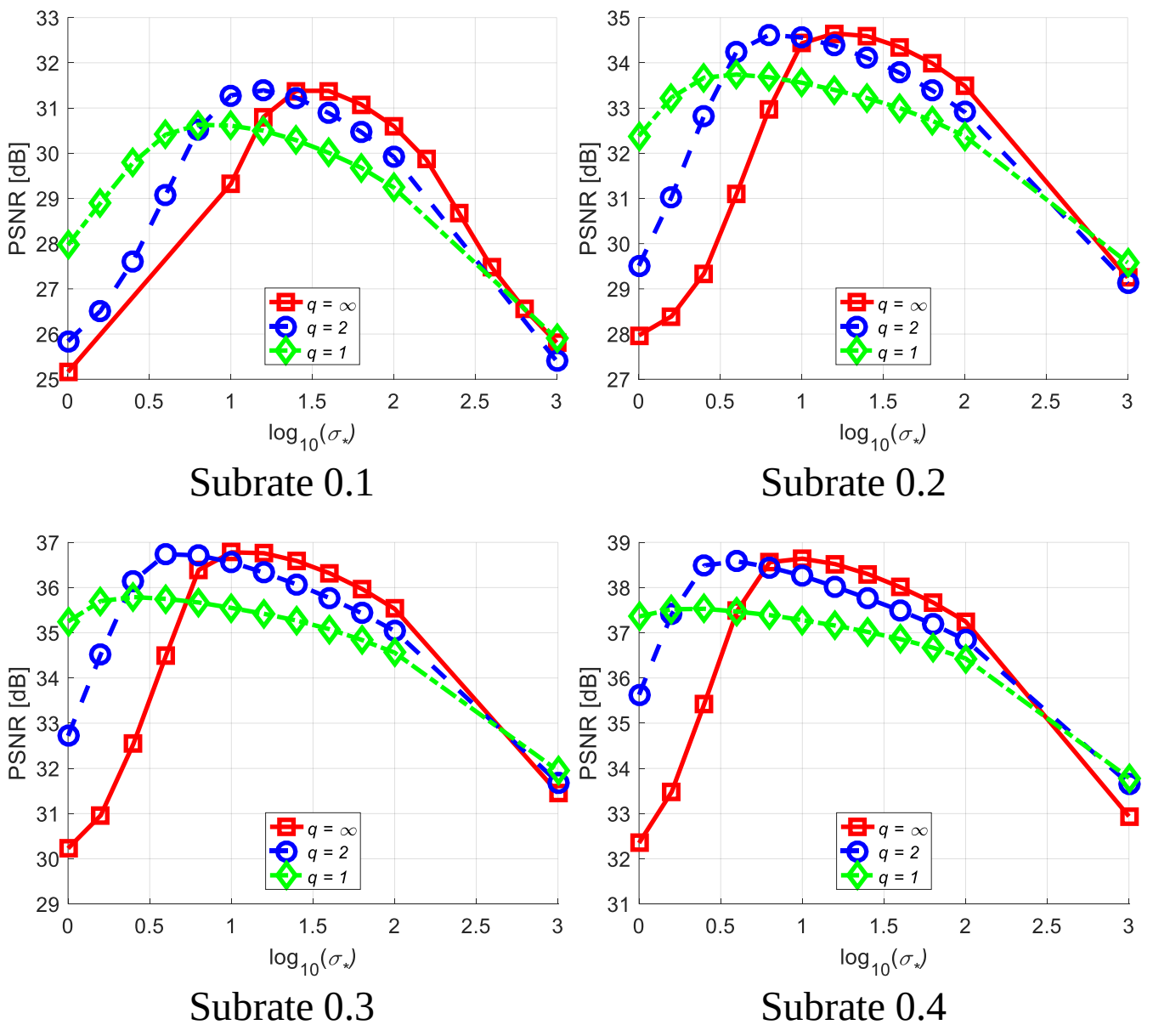

Figure 7. Average performance in PSNR [dB] of 8 test images for various $\sigma_*$ of the proposed HODW with $q$ = 1, 2, and ∞ with warm start by TVAL3 [11] and subrate = 0.1, 0.2, 0.3, and 0.4.

### A. *Performance*

#### 1) *Objective quality assessment*

The objective performance of the proposed HODW and other prior works are compared in PSNR and FSIMc [46] as in Table II and also summarized in Table III.

*Comparison of three cases of q*. Performance of the proposed HODW heavily depends on the tuning parameter $\sigma_*$, because it controls the strength of the filtering process through weighting in (24) and (29). Hence, we investigate the setting of $\sigma_*$ under three cases of $q$ = 1, 2, and ∞ in Fig. 7 which shows average performance of the proposed HODW over 8 test images with ranging from $10^0$ to $10^4$. Because more recovery error is expected in sparser representation of recovery images in case of low subrate $r$ = 0.1, $\sigma_*$ should be set larger for that case for stronger filtering of the sparse representation. On the other hand, the proposed HODW uses small $\sigma_*$ for the case of high subrate $r$ = 0.4 to do weak filtering. Therefore, as in Fig. 7, the peak performance is found at: $\sigma_* = \left\{10^{0.8}, 10^{0.6}, 10^{0.4}, 10^{0.4}\right\}$ for $q$ = 1, $\sigma_* = \left\{10^{1.2}, 10^{0.8}, 10^{0.6}, 10^{0.6}\right\}$ for $q$ = 2, and $\sigma_* = \left\{10^{1.6}, 10^{1.2}, 10^{1.0}, 10^{1.0}\right\}$ for $q$ = ∞ for subrates $r$ = 0.1, 0.2, 0.3, and 0.4, respectively.

Furthermore, by considering the best setting of $\sigma_*$ in our experiment for three case of $q$ = 1, 2, ∞, Fig. 7 shows that regardless of setting of subrate, HODW with $q$ = 1 (i.e., soft-

TABLE II. PERFORMANCE OF VARIOUS CS RECOVERY METHODS IN PSNR AND FSIMC [46] FOR COLOR IMAGES. THE AVERAGE AND THE BEST PERFORMANCE ARE IN ITALIC AND BOLD, RESPECTIVELY (THE BEST PERFORMANCE DOES NOT COUNT FOR $\mathrm{HODW}_{\mathrm{ORACLE}}$ BECAUSE IT IS NOT ACHIEVABLE IN PRACTICE).

| Quality index | PSNR [dB] | | | | | FSIMc | | | | | PSNR [dB] | | | | | FSIMc | | | | |
|---|---|---|---|---|---|---|---|---|---|---|---|---|---|---|---|---|---|---|---|---|
| Subrate | 0.1 | 0.2 | 0.3 | 0.4 | *Avg.* | 0.1 | 0.2 | 0.3 | 0.4 | *Avg.* | 0.1 | 0.2 | 0.3 | 0.4 | *Avg.* | 0.1 | 0.2 | 0.3 | 0.4 | *Avg.* |
| Image | Lena | | | | | | | | | | Goldhill | | | | | | | | | |
| $\ell_1$ [1] | 19.76 | 21.19 | 24.52 | 26.06 | 22.88 | 0.719 | 0.781 | 0.843 | 0.883 | 0.807 | 20.78 | 23.29 | 24.99 | 26.40 | 23.86 | 0.740 | 0.808 | 0.855 | 0.892 | 0.824 |
| SPL [12] | 25.31 | 27.60 | 29.21 | 30.72 | 28.21 | 0.835 | 0.891 | 0.922 | 0.945 | 0.898 | 25.27 | 26.98 | 28.31 | 29.52 | 27.52 | 0.824 | 0.875 | 0.907 | 0.929 | 0.884 |
| TVAL3 [11] | 25.77 | 28.35 | 30.23 | 31.95 | 29.08 | 0.822 | 0.893 | 0.931 | 0.954 | 0.900 | 25.84 | 27.88 | 29.34 | 30.72 | 28.44 | 0.815 | 0.884 | 0.919 | 0.943 | 0.890 |
| FPD [29] | 25.47 | 28.77 | 30.79 | 32.29 | 29.33 | 0.841 | 0.918 | 0.949 | 0.966 | 0.919 | 25.24 | 27.75 | 29.37 | 30.71 | 28.26 | 0.830 | 0.901 | 0.933 | 0.952 | 0.904 |
| NLR [14] | 28.56 | 30.90 | 32.60 | 34.07 | 31.53 | 0.900 | 0.945 | 0.964 | 0.975 | 0.946 | 26.93 | 28.91 | 30.56 | 32.05 | 29.61 | 0.858 | 0.906 | 0.937 | 0.956 | 0.914 |
| GSR [15] | 27.99 | 31.40 | 33.63 | 35.26 | 32.07 | 0.904 | 0.952 | 0.970 | 0.979 | 0.951 | 26.03 | 29.01 | 31.24 | 32.91 | 29.80 | 0.855 | 0.914 | 0.944 | 0.962 | 0.919 |
| SL20 [17] | 22.68 | 27.13 | 29.53 | 31.39 | 27.68 | 0.796 | 0.893 | 0.932 | 0.954 | 0.894 | 23.03 | 26.42 | 28.22 | 29.84 | 26.88 | 0.806 | 0.880 | 0.915 | 0.939 | 0.885 |
| SGIHT [18] | 24.91 | 27.43 | 29.36 | 31.02 | 28.18 | 0.822 | 0.888 | 0.924 | 0.947 | 0.895 | 25.18 | 27.16 | 28.62 | 29.98 | 27.74 | 0.821 | 0.881 | 0.914 | 0.937 | 0.888 |
| $\mathrm{HODW}_{q=1}$ | 30.60 | 33.22 | 34.86 | 36.30 | 33.74 | 0.934 | 0.965 | 0.977 | 0.984 | 0.965 | 29.56 | 31.91 | 33.75 | 35.28 | 32.62 | 0.905 | 0.945 | 0.967 | 0.977 | 0.949 |
| $\mathrm{HODW}_{q=2}$ | 31.14 | 33.81 | **35.48** | 36.98 | 34.35 | **0.943** | **0.971** | **0.981** | **0.987** | **0.971** | **29.89** | **32.47** | **34.44** | **36.15** | **33.24** | **0.917** | **0.955** | **0.974** | **0.982** | **0.957** |
| $\mathrm{HODW}_{q=\infty}$ | **31.15** | **33.84** | 35.46 | **37.03** | **34.37** | 0.942 | **0.971** | **0.981** | **0.987** | 0.970 | 29.82 | 32.39 | 34.32 | 36.14 | 33.17 | 0.915 | **0.955** | 0.973 | **0.982** | 0.956 |
| $\mathrm{HODW}_{\mathrm{Oracle}}$ | 36.78 | 40.78 | 44.84 | 49.63 | 43.01 | 0.985 | 0.995 | 0.998 | 0.999 | 0.994 | 35.63 | 40.34 | 45.02 | 50.25 | 42.81 | 0.977 | 0.994 | 0.998 | 0.999 | 0.992 |
| Image | Barbara | | | | | | | | | | Boats | | | | | | | | | |
| $\ell_1$ [1] | 19.05 | 20.89 | 22.45 | 23.83 | 21.56 | 0.681 | 0.753 | 0.813 | 0.856 | 0.776 | 18.69 | 21.31 | 23.74 | 25.41 | 22.29 | 0.695 | 0.772 | 0.842 | 0.878 | 0.797 |
| SPL [12] | 21.23 | 22.24 | 23.20 | 24.48 | 22.78 | 0.753 | 0.798 | 0.836 | 0.873 | 0.815 | 24.66 | 27.51 | 29.65 | 31.40 | 28.30 | 0.831 | 0.894 | 0.929 | 0.951 | 0.901 |
| TVAL3 [11] | 21.09 | 22.78 | 24.80 | 27.20 | 23.97 | 0.743 | 0.824 | 0.885 | 0.930 | 0.846 | 25.55 | 28.93 | 31.64 | 33.93 | 30.01 | 0.848 | 0.920 | 0.954 | 0.972 | 0.924 |
| FPD [29] | 21.58 | 24.50 | 27.37 | 30.07 | 25.88 | 0.769 | 0.868 | 0.924 | 0.956 | 0.879 | 24.91 | 29.83 | 33.28 | 35.75 | 30.94 | 0.839 | 0.938 | 0.970 | 0.984 | 0.933 |
| NLR [14] | 25.51 | 29.44 | 31.04 | 32.90 | 29.72 | 0.892 | 0.952 | 0.968 | 0.979 | 0.948 | 30.49 | 35.17 | 37.34 | 38.65 | 35.41 | 0.940 | 0.978 | 0.987 | 0.991 | 0.974 |
| GSR [15] | 28.72 | 33.37 | 35.72 | 37.52 | 33.83 | 0.944 | 0.977 | 0.986 | 0.991 | 0.975 | 30.69 | 35.36 | 37.90 | 39.69 | 35.91 | 0.946 | 0.979 | 0.988 | 0.992 | 0.976 |
| SL20 [17] | 21.36 | 25.60 | 28.67 | 31.46 | 26.77 | 0.779 | 0.885 | 0.933 | 0.962 | 0.890 | 22.34 | 27.87 | 31.62 | 34.50 | 29.08 | 0.789 | 0.904 | 0.950 | 0.974 | 0.904 |
| SGIHT [18] | 21.19 | 22.14 | 23.14 | 25.11 | 22.90 | 0.749 | 0.798 | 0.838 | 0.888 | 0.818 | 24.41 | 27.49 | 29.70 | 31.62 | 28.30 | 0.822 | 0.896 | 0.932 | 0.955 | 0.901 |
| $\mathrm{HODW}_{q=1}$ | 31.14 | 34.98 | 37.03 | 38.77 | 35.48 | 0.958 | 0.982 | 0.989 | 0.993 | 0.981 | 33.74 | 37.38 | 39.60 | 41.24 | 37.99 | 0.966 | 0.986 | 0.992 | 0.995 | 0.985 |
| $\mathrm{HODW}_{q=2}$ | 32.60 | 35.94 | 37.95 | 39.67 | 36.54 | **0.969** | 0.985 | **0.991** | 0.994 | **0.985** | **34.83** | 38.46 | 40.47 | 42.14 | **38.98** | **0.973** | **0.989** | **0.993** | **0.996** | **0.988** |
| $\mathrm{HODW}_{q=\infty}$ | **32.71** | **36.05** | **38.03** | **39.77** | **36.64** | **0.969** | **0.986** | **0.991** | **0.995** | **0.985** | 34.73 | **38.48** | **40.49** | **42.19** | 38.97 | 0.972 | **0.989** | **0.993** | **0.996** | **0.988** |
| $\mathrm{HODW}_{\mathrm{Oracle}}$ | 39.55 | 44.04 | 48.51 | 53.54 | 46.41 | 0.994 | 0.998 | 0.999 | 1.000 | 0.998 | 42.02 | 46.46 | 51.15 | 56.38 | 49.01 | 0.995 | 0.998 | 1.000 | 1.000 | 0.998 |
| Image | Peppers | | | | | | | | | | House | | | | | | | | | |
| $\ell_1$ [1] | 18.07 | 21.31 | 23.58 | 25.31 | 22.07 | 0.727 | 0.786 | 0.833 | 0.866 | 0.803 | 18.54 | 20.17 | 23.40 | 25.04 | 21.79 | 0.711 | 0.760 | 0.843 | 0.883 | 0.799 |
| SPL [12] | 25.18 | 28.60 | 31.05 | 33.13 | 29.49 | 0.861 | 0.914 | 0.944 | 0.961 | 0.920 | 23.61 | 26.01 | 27.71 | 29.28 | 26.65 | 0.814 | 0.874 | 0.908 | 0.933 | 0.882 |
| TVAL3 [11] | 26.63 | 30.78 | 33.68 | 36.18 | 31.82 | 0.886 | 0.942 | 0.965 | 0.979 | 0.943 | 24.38 | 27.30 | 29.42 | 31.33 | 28.11 | 0.811 | 0.891 | 0.931 | 0.956 | 0.897 |
| FPD [29] | 25.65 | 30.89 | 34.08 | 36.21 | 31.71 | 0.862 | 0.942 | 0.969 | 0.980 | 0.938 | 23.75 | 27.53 | 30.28 | 32.58 | 28.53 | 0.819 | 0.907 | 0.945 | 0.966 | 0.909 |
| NLR [14] | 31.87 | 36.14 | 38.28 | 40.07 | 36.59 | 0.952 | 0.978 | 0.986 | 0.992 | 0.977 | 26.78 | 30.63 | 33.27 | 35.58 | 31.56 | 0.885 | 0.948 | 0.972 | 0.984 | 0.947 |
| GSR [15] | 31.73 | 36.07 | 38.51 | 40.32 | 36.66 | 0.955 | 0.978 | 0.987 | 0.991 | 0.978 | 26.37 | 30.89 | 34.11 | 36.60 | 31.99 | 0.894 | 0.953 | 0.975 | 0.986 | 0.952 |
| SL20 [17] | 21.09 | 26.14 | 29.39 | 32.00 | 27.16 | 0.756 | 0.857 | 0.912 | 0.944 | 0.867 | 21.35 | 25.93 | 28.82 | 31.23 | 26.83 | 0.790 | 0.885 | 0.929 | 0.956 | 0.890 |
| SGIHT [18] | 24.68 | 27.97 | 30.52 | 32.78 | 28.99 | 0.848 | 0.907 | 0.939 | 0.959 | 0.913 | 23.50 | 26.13 | 28.07 | 29.81 | 26.88 | 0.812 | 0.880 | 0.916 | 0.941 | 0.887 |
| $\mathrm{HODW}_{q=1}$ | 31.87 | 35.79 | 38.28 | 40.19 | 36.53 | 0.955 | 0.978 | 0.987 | 0.992 | 0.978 | 29.26 | 32.89 | 35.40 | 37.62 | 33.79 | 0.928 | 0.966 | 0.981 | 0.989 | 0.966 |
| $\mathrm{HODW}_{q=2}$ | 33.04 | 37.12 | 39.57 | 41.40 | 37.78 | 0.962 | **0.983** | **0.990** | **0.994** | 0.982 | 30.03 | 34.13 | 36.90 | 39.26 | 35.08 | 0.941 | 0.975 | 0.987 | **0.993** | 0.974 |
| $\mathrm{HODW}_{q=\infty}$ | **33.18** | **37.23** | **39.66** | **41.47** | **37.89** | **0.963** | **0.983** | **0.990** | **0.994** | **0.983** | **30.12** | **34.28** | **37.08** | **39.43** | **35.22** | **0.943** | **0.976** | **0.988** | **0.993** | **0.975** |
| $\mathrm{HODW}_{\mathrm{Oracle}}$ | 41.09 | 45.72 | 49.74 | 54.19 | 47.68 | 0.993 | 0.998 | 0.999 | 1.000 | 0.998 | 38.44 | 44.33 | 48.87 | 53.27 | 46.23 | 0.991 | 0.998 | 0.999 | 1.000 | 0.997 |
| Image | Mandrill | | | | | | | | | | Brick | | | | | | | | | |
| $\ell_1$ [1] | 19.02 | 20.84 | 22.13 | 23.20 | 21.30 | 0.648 | 0.733 | 0.799 | 0.843 | 0.756 | 20.25 | 22.95 | 25.58 | 27.56 | 24.08 | 0.733 | 0.804 | 0.859 | 0.897 | 0.823 |
| SPL [12] | 21.78 | 23.10 | 24.17 | 25.22 | 23.56 | 0.732 | 0.803 | 0.851 | 0.885 | 0.818 | 26.56 | 29.53 | 31.59 | 33.17 | 30.21 | 0.848 | 0.902 | 0.931 | 0.951 | 0.908 |
| TVAL3 [11] | 21.65 | 23.03 | 24.43 | 25.80 | 23.73 | 0.699 | 0.796 | 0.857 | 0.896 | 0.812 | 29.19 | 32.32 | 34.33 | 35.80 | 32.91 | 0.886 | 0.933 | 0.957 | 0.970 | 0.937 |
| FPD [29] | 21.49 | 23.28 | 24.73 | 26.00 | 23.88 | 0.747 | 0.832 | 0.886 | 0.915 | 0.845 | 28.78 | 32.48 | 34.38 | 35.86 | 32.87 | 0.895 | 0.950 | 0.969 | 0.979 | 0.948 |
| NLR [14] | 21.70 | 24.06 | 25.88 | 27.33 | 24.74 | 0.738 | 0.847 | 0.896 | 0.923 | 0.851 | 33.34 | 35.56 | 37.32 | 39.05 | 36.32 | 0.937 | 0.974 | 0.986 | 0.991 | 0.972 |
| GSR [15] | 20.20 | 22.99 | 25.46 | 27.83 | 24.12 | 0.740 | 0.856 | 0.906 | 0.939 | 0.860 | 33.03 | 36.26 | 38.04 | 39.53 | 36.71 | 0.939 | 0.974 | 0.985 | 0.990 | 0.972 |
| SL20 [17] | 19.94 | 22.23 | 23.69 | 25.22 | 22.77 | 0.743 | 0.823 | 0.870 | 0.905 | 0.835 | 24.28 | 30.19 | 32.86 | 34.87 | 30.55 | 0.805 | 0.921 | 0.956 | 0.974 | 0.914 |
| SGIHT [18] | 21.69 | 23.23 | 24.45 | 25.65 | 23.75 | 0.723 | 0.807 | 0.858 | 0.893 | 0.820 | 25.79 | 28.60 | 30.80 | 32.65 | 29.46 | 0.825 | 0.887 | 0.920 | 0.946 | 0.895 |
| $\mathrm{HODW}_{q=1}$ | 24.21 | 26.88 | 28.98 | 31.04 | 27.78 | 0.828 | 0.905 | 0.942 | 0.965 | 0.910 | 34.69 | 36.86 | 38.41 | 39.79 | 37.44 | 0.960 | 0.978 | 0.987 | 0.992 | 0.979 |
| $\mathrm{HODW}_{q=2}$ | **24.51** | **27.68** | 30.14 | 32.61 | **28.73** | 0.864 | 0.928 | 0.959 | 0.976 | 0.932 | **35.08** | **37.38** | 38.94 | 40.40 | **37.95** | **0.964** | **0.983** | **0.990** | **0.993** | **0.983** |
| $\mathrm{HODW}_{q=\infty}$ | 24.44 | 27.57 | **30.28** | **32.65** | **28.73** | **0.867** | **0.929** | **0.961** | **0.977** | **0.934** | 34.92 | 37.37 | **38.97** | **40.42** | 37.92 | 0.963 | **0.983** | **0.990** | **0.993** | 0.982 |
| $\mathrm{HODW}_{\mathrm{Oracle}}$ | 31.89 | 37.65 | 42.55 | 47.40 | 39.87 | 0.967 | 0.992 | 0.998 | 0.999 | 0.989 | 40.35 | 44.82 | 49.39 | 53.94 | 47.12 | 0.991 | 0.997 | 0.999 | 1.000 | 0.997 |

thresholding) is less effective than $q = 2$ (i.e., weighted soft-thresholding) and $q = \infty$ (i.e., hard-thresholding). This is because the weight tensor $W$ with a larger $q$ (that is, 2 or $\infty$) promotes sparsity of the recovered sparse representation more than $q = 1$, as interpreted from (29). Therefore $\mathrm{HODW}_{q=\infty}$ is 0.94 dB better than $\mathrm{HODW}_{q=1}$ and slightly (0.03 dB) better than $\mathrm{HODW}_{q=2}$.

*Oracle performance of HODW.* To illustrate the discussion on design of weight tensor $W$ from filtering viewpoint with the minimum mean squared error approach, we further investigate the $\mathrm{HODW}_{\mathrm{Oracle}}$. It significantly improves the quality of the recovered images, for example of Lena in Table II, it is 5.63 dB and 12.60 dB better than the $\mathrm{HODW}_{q=\infty}$ at subrate 0.1 and 0.4, respectively. The gains of 5.81 dB and 14.11 dB are found for Goldhill at the same conditions. Average performance of 8 test images in Table III confirms that the improvements by $\mathrm{HODW}_{\mathrm{Oracle}}$ increases as subrate increases. The improvement is 6.83 dB at subrate 0.1 and is up to 13.69 dB at subrate 0.4. Though the $\mathrm{HODW}_{\mathrm{Oracle}}$ is not practical, it does show that there is still plenty of room to improve for the proposed HODW by a better design of the weight tensor $W$.

TABLE III. DELTA PSNR [DB] OF THE PROPOSED $\text{HODW}_{Q=\infty}$ FROM OTHER RECOVERIES, A POSITIVE NUMBER INDICATES BETTER PERFORMANCE OF THE PROPOSED $\text{HODW}_{Q=\infty}$.

| Subrate | 0.1 | 0.2 | 0.3 | 0.4 | Average |
|---|---|---|---|---|---|
| $\ell_1$ [1] | 12.11 | 13.16 | 12.99 | 13.29 | 12.89 |
| SPL [12] | 7.18 | 8.21 | 8.68 | 9.02 | 8.27 |
| TVAL3 [11] | 6.37 | 6.98 | 7.05 | 7.02 | 6.86 |
| FPD [29] | 6.77 | 6.52 | 6.25 | 6.20 | 6.44 |
| NLR [14] | 3.23 | 3.30 | 3.50 | 3.67 | 3.43 |
| GSR [15] | 3.29 | 2.73 | 2.46 | 2.43 | 2.73 |
| SL20 [17] | 9.37 | 8.21 | 7.68 | 7.32 | 8.15 |
| SGIHT [18] | 7.46 | 8.38 | 8.70 | 8.81 | 8.34 |
| $\text{HODW}_{q=1}$ | 0.75 | 0.91 | 1.00 | 1.11 | 0.94 |
| $\text{HODW}_{q=2}$ | -0.01 | 0.03 | 0.05 | 0.06 | 0.03 |
| $\text{HODW}_{q=\infty}$ | - | - | - | - | - |
| $\text{HODW}_{\text{Oracle}}$ | -6.83 | -8.37 | -10.72 | -13.69 | -9.90 |

*Comparison to other state-of-the-art methods*. By exploiting the nonlocal property of color images, the proposed HODW is significantly better than the previous work employing the fixed sparsifying bases of DCT (in $\ell_1$ [1]), DWT (in SPL [12]), gradient (in TVAL3 [11]), and even trained dictionary (in FPD [29]) regardless of their sparsity regularizations. For an example of subrate 0.1 for Lena in Table II, their performances in PSNR (and FSIMc) are 19.76 dB (0.719), 25.31 dB (0.835), 25.77 dB (0.822), and 25.47 dB (0.841), respectively, whereas $\text{HODW}_{q=\infty}$ has performance of 31.15 dB (0.942). The same comparison is also applied for other images in other subrates. By averaging all test cases, $\text{HODW}_{q=\infty}$ is shown to achieve performance improvement by more than 6 dB to the methods of $\ell_1$ [1], SPL [12], and TVAL3 [11] as shown in Table III.

By exploiting the nonlocal property of images, NLR [14] and GSR [15] can produce better performance than the recoveries with fixed sparsifying bases. However, without taking the color correlation in account, as depicted in Fig. 2, their performance is inferior to the proposed HODW that utilizes the color correlation. For the image of Brick which has less color correlation as shown in Fig. 2, the gain by $\text{HODW}_{q=\infty}$ is only 1.89 dB compared to GSR [15] at subrate 0.1 as interpreted from Table II. However, for Boats having high color correlation, $\text{HODW}_{q=\infty}$ can give improvement up to 4.04 dB compared to GSR [15] at subrate 0.1. On average in Table III, $\text{HODW}_{q=\infty}$ is by 2.73 dB and 3.43 dB better than GSR [15] and NLR [14], respectively.

Finally, the proposed HODW is also significantly better than the other recovery methods for color images, SL20 [17] and SGIHT [18]. This is because, instead of using $\ell_{20}$ as in SL20 [17] and SGIHT [18], HODW utilizes HOSVD to exploit the color correlation, and then, at the same time the nonlocal property of color images as well. The performances at subrate 0.1 for Lena of SL20 [17], SGIHT [18], and $\text{HODW}_{q=\infty}$ are respectively 22.68 dB (0.796), 24.92 dB (0.822), and 31.15 dB (0.942). On average, in Table III, SL20 [17] and SGIHT's [18] performance are by 8.15 dB and 8.34 dB inferior to $\text{HODW}_{q=\infty}$.

*2) Subjective quality assessment*

Subjective quality is consistent to objective quality of PSNR or FSIMc [46] in Table II, where the proposed HODW brings significant improvement to the state-of-the-art methods. We evaluate the subjective quality of the recovered images for Lena, Mandrill, and Goldhill by different recovery methods with respect to various sparsifying bases and various sparsity regularizations as shown in Fig. 1, 8, and 9. By using the fixed sparsifying bases of DCT or wavelet transform or trained dictionary $\ell_1$ [1], SPL [12], TVAL3 [11], FPD [29], most of texture information cannot be recovered, for example of texture in cheek of mandrill in Fig. 8 or the window in box in Fig. 9. With the fixed sparsifying bases, SL20 [17] and SGIHT [18] which are even dedicated to color images still cannot recover those texture information well, for example, the window in box in Fig. 9 is blurred. On the other hand, with nonlocal exploitation, this texture information is recovered better by NLR [14] and GSR [15]. However, the works of NLR [14] and GSR [15], without color exploitation, have problem of color mismatch among three color channels, and this produces weird color appearance, for example, nearby the mouth of mandrill in Fig. 8 or the mouth and the left eye of Lena in Fig. 1. This problem is solved by recovery of three color channels together by the proposed HODW, by which the weird color disappears. Furthermore, color exploitation also helps preserving texture information better, as seen in the hat of Lena, mandrill's hair, and the window recovered by proposed HODW in Fig.'s 1, 8, and 9.

## B. *Performance evolution*

The convergence of the proposed method is not easy to prove due to some approximations in (16) and (27) and updates by each iteration for the dictionary $D$ and weight tensor $W$. Hence, in this paper, we experimentally show the performance evolution in PSNR of the proposed HODW for both cases, with and without warm start in Fig. 10 for several test images. Fig. 10 shows that the proposed HODW evolves to stable states of PSNR for both cases of with and without warm start, however, it evolves faster with warm start (around 50 iterations) than without warm start (around 200 iterations).

## C. *Computational complexity*

Compared to those frameworks based on a fixed sparsifying bases of DCT in $\ell_1$-DCT [1], DWT in SPL-DWT [12], and gradient in TVAL3 [11], those frameworks based on the nonlocal property of GSR [15] and NLR [14] have to perform computationally heavy tasks of searching for similar patches and singular value decomposition. Therefore, they consume more computational time than recoveries with fixed sparsifying bases. GSR [15], NLR [14], and the proposed HODW consume 7433*s*, 1369*s*, and 5704*s*, respectively as shown in Table IV. However, with cost of more computation, the frameworks based on the nonlocal property of images bring dramatic improvement to CS applications of images.

# VII. CONCLUSION AND FUTURE WORK

This paper proposed a learning scheme of an adaptive higher-order dictionary that uses higher order singular value decomposition to exploit correlation within color images in compressive sensing framework. Its corresponding sparse representation is recovered with a weighted sparsity regularization accompanied with a weight tensor to promote the sparsity of the recovered sparse representation, so that the large sparse representation's elements (in magnitude) are expected to be recovered as large values and vice versa. By this concept, we design the weight tensor from the viewpoint of minimum mean squared error filtering (i.e., Wiener filter) with several variations of the designs of the weight tensor. Our experimental results confirmed its performance improvement over other state-of-the-art recovery methods.

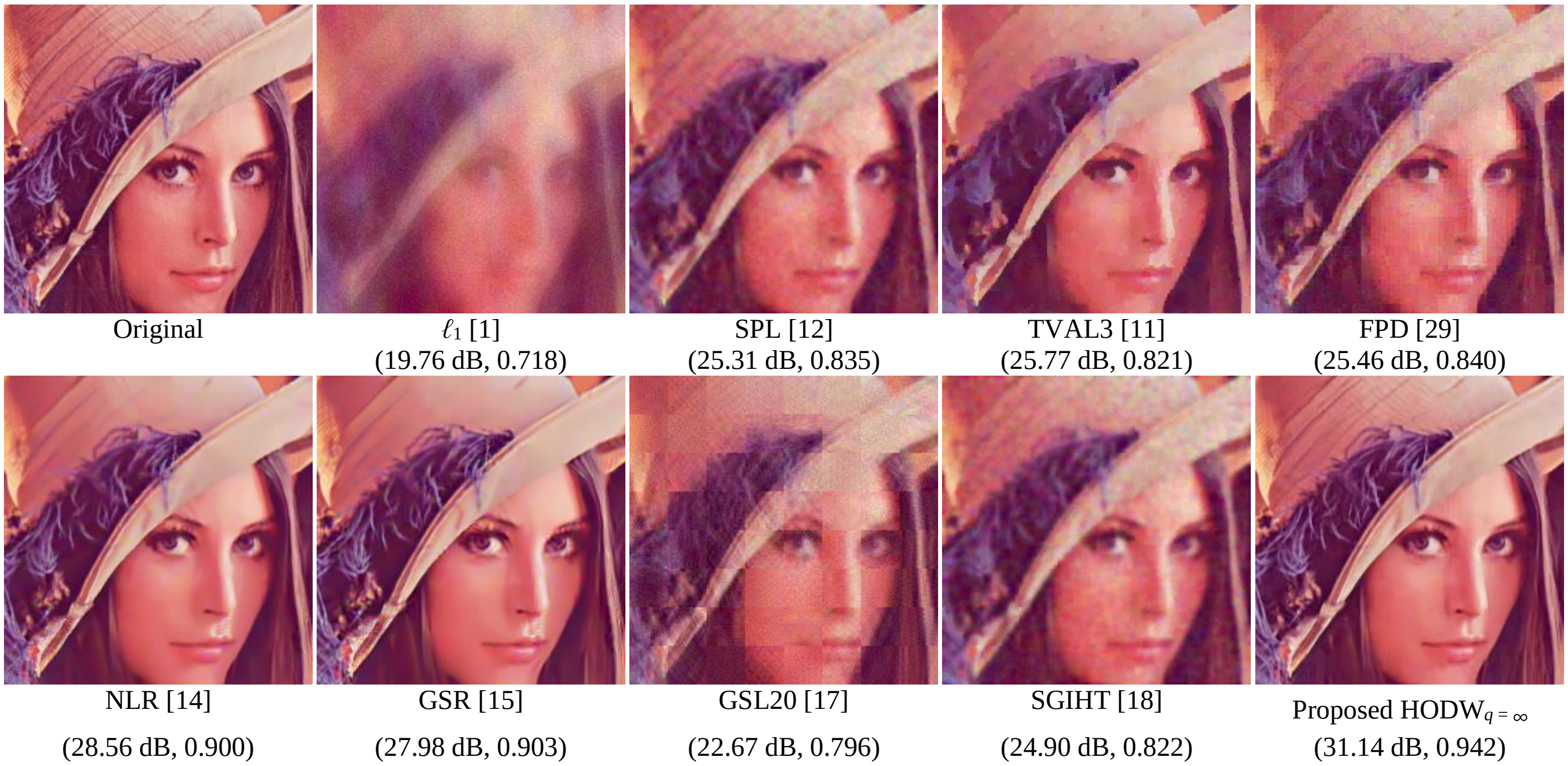

Figure 8. Comparison of recovered color image Lena by different recovery methods employing different sparsity bases and sparsity regularizations. Numbers in brackets are PSNR and FSIMc [46], respectively. It uses a structural sensing matrix [45] at subrate 0.1.

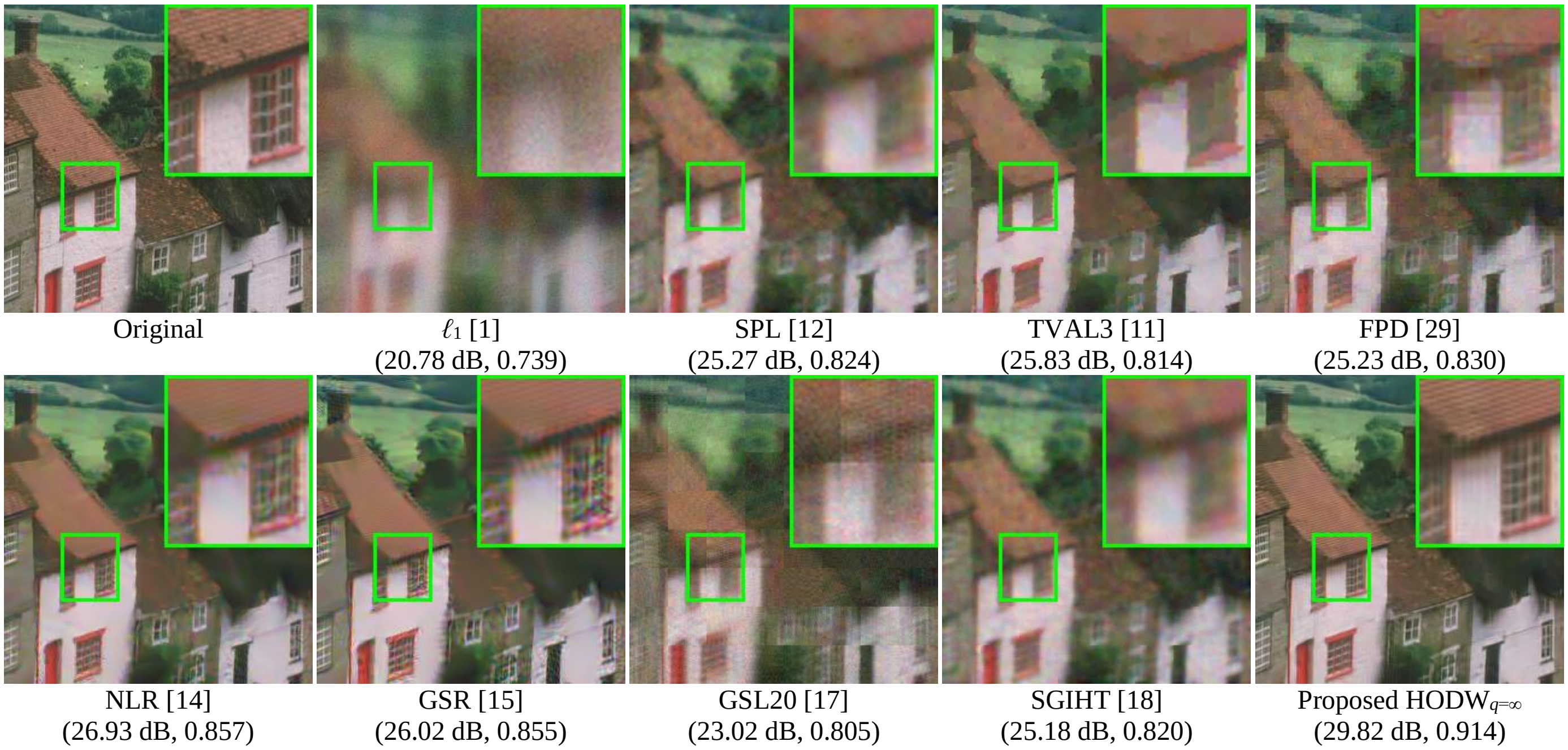

Figure 9. Comparison of recovered color image Goldhill by different recovery methods employing different sparsity bases and sparsity regularizations. Numbers in brackets are PSNR and FSIMc [46], respectively. It uses a structural sensing matrix [45] at subrate 0.1.

Furthermore, the proposed method supplied by optimal weighted tensor as an oracle case shows possibility of dramatic improvement, which sheds lights on further development of better design of the weight tensor to approximate the oracle case. Besides, the proposed method is solved for general case of the sensing matrix $\Phi$, which is easily extended to other image restoration problems, for example, deblurring of color images, where $\Phi$ is the corresponding blurring kernels. Finally, this paper showed the case of compressive sensing, in which the dictionary and the weight tensor are learned from the recovery of previous iteration. However, it can be extended to other works in which the reference picture is available such as the case of flash/no flash denoising.

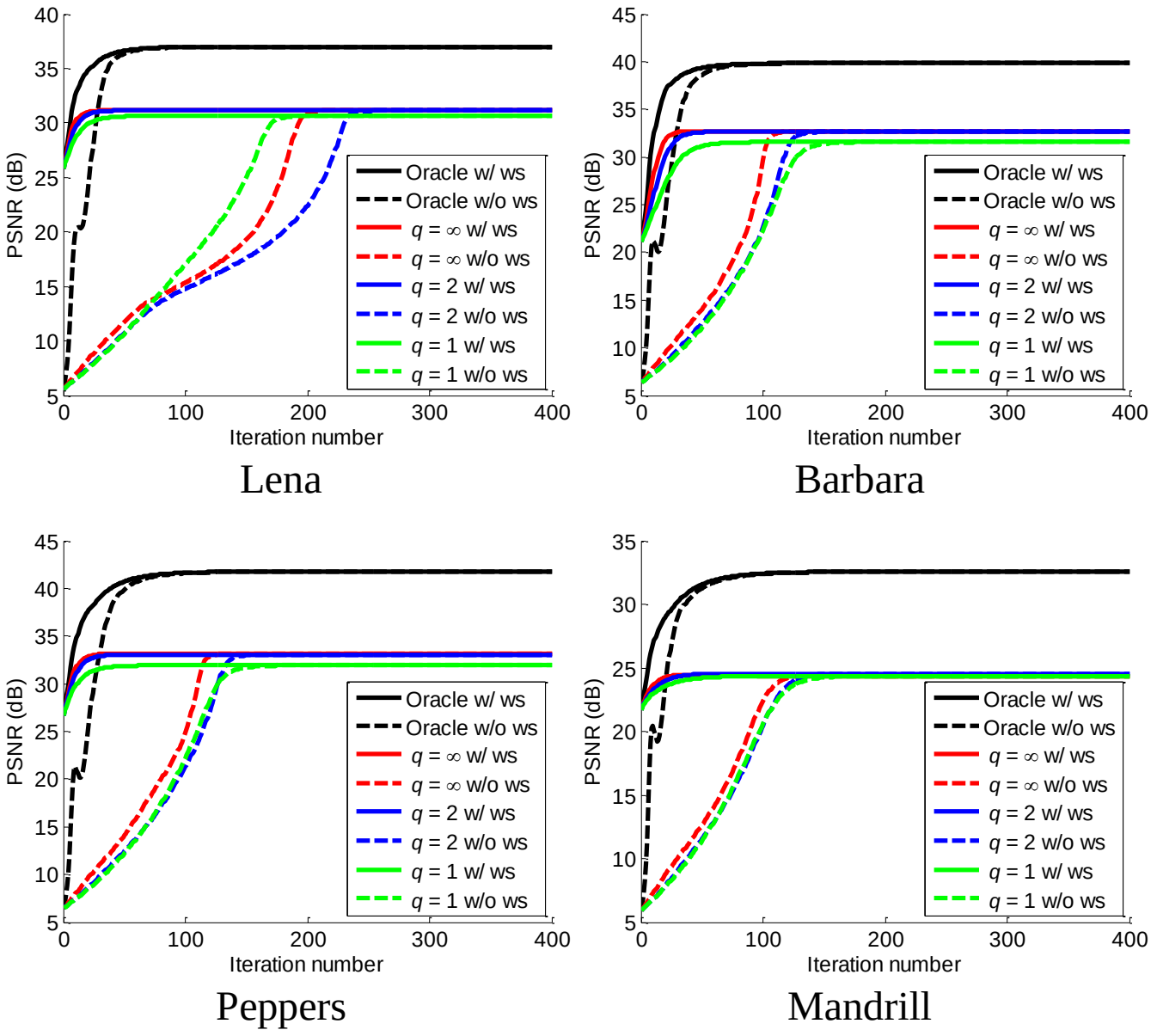

Figure 10. Performance evolution in PSNR of the proposed HODW with and without warm start of TVAL3 [11] (denote "w/ ws" and "w/o ws", respectively) at subrate 0.1.

TABLE IV
AVERAGE RECOVERY TIME (IN SECONDS) FOR ALL TEST IMAGES.

| Subrate | 0.1 | 0.2 | 0.3 | 0.4 | *Avg.* |
|---|---|---|---|---|---|
| $\ell_1$ [1] | 88 | 83 | 111 | 101 | 96 |
| SPL [12] | 17 | 8 | 5 | 4 | 8 |
| TVAL3 [11] | 10 | 8 | 8 | 7 | 8 |
| FPD [29] | 501 | 508 | 512 | 514 | 509 |
| NLR [14] | 1316 | 1352 | 1388 | 1421 | 1369 |
| GSR [15] | 7375 | 7384 | 7461 | 7513 | 7433 |
| SL20 [17] | 42 | 43 | 44 | 43 | 43 |
| SGIHT [18] | 40 | 21 | 15 | 10 | 22 |
| $\text{HODW}_{q=\infty}$ | 5704 | 5704 | 5700 | 5706 | 5704 |

## ACKNOWLEDGEMENT

This work was supported by MSIP through the National Research Foundation of Korea (Grant 2011-001-7578) and G-ITRC support program (IITP-2016-R6812-16-0001) supervised by the IITP.